%% file: conference_101719.tex
\documentclass[conference]{IEEEtran}
\IEEEoverridecommandlockouts
\usepackage{cite}
\usepackage{tikz}
\usepackage{multirow}
\usepackage{listings}
\usepackage{xcolor}
\usepackage{tablefootnote}
\definecolor{codegreen}{rgb}{0,0.6,0}
\definecolor{codegray}{rgb}{0.5,0.5,0.5}
\definecolor{codepurple}{rgb}{0.58,0,0.82}
\definecolor{backcolour}{rgb}{0.95,0.95,0.92}
\usepackage{url}
\lstdefinestyle{mystyle}{
    backgroundcolor=\color{backcolour},   
    commentstyle=\color{codegreen},
    keywordstyle=\color{magenta},
    numberstyle=\tiny\color{codegray},
    stringstyle=\color{codepurple},
    basicstyle=\ttfamily\footnotesize,
    breakatwhitespace=false,         
    breaklines=true,                 
    captionpos=b,                    
    keepspaces=true,                 
    numbers=left,                    
    numbersep=5pt,                  
    showspaces=false,                
    showstringspaces=false,
    showtabs=false,                  
    tabsize=2
}

\usetikzlibrary{shapes.geometric}
\usetikzlibrary{arrows.meta,arrows,positioning}
\usepackage{amsmath,amssymb,amsfonts}
\usepackage{algorithm2e}                
\usepackage{enumitem}                   
\usepackage{mathtools}                  
\usepackage{microtype}                  
\usepackage{float}    

\usepackage{algpseudocode}%
\usepackage{graphicx}
\usepackage{textcomp}
\usepackage{xcolor}
\usepackage{algorithmicx}

\usepackage{enumitem}
\def\BibTeX{{\rm B\kern-.05em{\sc i\kern-.025em b}\kern-.08em
    T\kern-.1667em\lower.7ex\hbox{E}\kern-.125emX}}

\usepackage{xcolor}

\begin{document}

\title{Measuring LLM Code Generation Stability 
via Structural Entropy}
\author{
 \textbf{Yewei Song,}
 \textbf{Tiezhu Sun,}
 \textbf{Xunzhu Tang,}
 \textbf{Prateek Kumar Rajput,}
 \\
 \textbf{Tegawendé F. Bissyandé}
 \textbf{Jacques Klein,}
\\The Interdisciplinary Centre for Security, Reliability and Trust\\
 University of Luxembourg
 
}

\pagestyle{plain} 
\maketitle

\begin{abstract}
Assessing the stability of code generation from large language models (LLMs) is essential for judging their reliability in real-world development. We extend prior “structural‐entropy’’ concepts to the program domain by pairing entropy with abstract-syntax-tree (AST) analysis. For any fixed prompt, we collect the multiset of depth-bounded subtrees of AST in each generated program and treat their relative frequencies as a probability distribution. We then measure stability in two complementary ways: (i) \emph{Jensen–Shannon divergence}, a symmetric, bounded indicator of structural overlap, and (ii) a \emph{Structural Cross-Entropy} ratio that highlights missing high-probability patterns. Both metrics admit structural-only and token-aware variants, enabling separate views on control-flow shape and identifier-level variability. Unlike pass@$k$, BLEU, or CodeBLEU, our metrics are reference-free, language-agnostic, and execution-independent. We benchmark several leading LLMs on standard code generation tasks, demonstrating that AST-driven structural entropy reveals nuances in model consistency and robustness. The method runs in $O(n,d)$ time with no external tests, providing a lightweight addition to the code-generation evaluation toolkit.
\end{abstract}


\section{Introduction}
The advent of powerful Large Language Models (LLMs) has enabled remarkable capabilities in automated code generation. However, a notable challenge is the high variability of generated code: identical prompts can yield substantially different code snippets across runs or across models. Prior studies observe that even with fixed inputs and hyperparameters, state-of-the-art LLMs are rarely deterministic at the output level~\cite{atil2024llm}. For example, ChatGPT produces completely different code on repeated queries for the same programming task in the majority of cases (e.g. 75.8\% of tasks showed zero identical test outputs across runs)~\cite{ouyang2025empirical}. Crucially, setting the sampling temperature to zero (greedy decoding) did not guarantee consistency. This output instability undermines developer trust and makes reproducibility of code-generation research difficult. In safety-critical or collaborative software settings, unpredictable LLM suggestions can harm reliability. Hence there is a pressing need to rigorously quantify the structural stability of generated code, beyond simply assessing correctness.

Current evaluation of LLM code outputs largely focuses on functional correctness or textual similarity to reference solutions. The pass@k metric, for example, checks if any of k samples passes the unit tests~\cite{chen2021evaluating}, thereby measuring functional success. 
While pass@k and related metrics (e.g. the unbiased pass-ratio@n) are valuable for overall performance~\cite{jiang2024survey}, they do not address variability: different outputs may all pass the tests but differ substantially in structure.
Likewise, traditional NLP-style metrics (BLEU, ROUGE, METEOR, etc.~\cite{papineni2002bleu,banerjee-lavie-2005-meteor,lin2004rouge}) assess n-gram overlap with a reference solution, but have known limitations for code~\cite{evtikhiev2023out,tran2019does}. Ren et al. in the CodeBLEU work show that BLEU correlates poorly with code semantics and cannot account for the many functionally equivalent programs that do not share surface tokens\cite{ren2020codebleu}. Code-specific metrics such as CodeBLEU have been proposed to address syntax and semantics: CodeBLEU augments n-gram matches with AST-based syntax weighting and data-flow features~\cite{ren2020codebleu}. These improvements help correlate with human judgments, but CodeBLEU and similar metrics still measure pairwise similarity to a single reference output rather than the consistency across multiple samples. We list some metrics that are currently used for measuring LLM stability in Table \ref{tab:compare}. In short, existing metrics focus on correctness or reference fidelity, but they do not capture structural/topological similarity of different outputs from the same prompt.

Although structural entropy—measuring uncertainty or variability in structural choices—is not widely used in LLM code-generation literature, related concepts have been explored in software engineering contexts. For example, Torres et al. applied structural entropy metrics to study software evolution, quantifying how changes affect the organization and complexity of code structures over time~\cite{torres2023applying}. Their method captures how transformations impact predictability and structural information in software systems. A closely analogous idea in LLM text generation is Semantic Entropy (SE), proposed by Kossen et al., which assesses model uncertainty by clustering multiple generated answers by meaning; high SE indicates diverse or inconsistent outputs, signaling possible hallucinations~\cite{kossen2024semantic}. Inspired by this, our work adapts the entropy concept to code generation, analyzing variability in generated code AST structures. Low structural entropy reflects consistent outputs, whereas high entropy indicates the model frequently alternates between different programming structures or approaches.
\input{table_compare}
\input{approaches}

\input{exp}

\section{Threats to Validity}

Several threats may impact the validity of our findings. Firstly, our experiments were conducted on specific benchmarks (Python BigCodeBench and SQL Spider), potentially limiting generalizability across other languages and tasks. Second, the chosen depth parameter in subtree extraction might influence stability measures; different values could yield varied results. Additionally, while our entropy-based metrics effectively capture syntactic variability, they currently do not explicitly address semantic equivalences or execution behaviors. Lastly, the limited number of models evaluated and the constrained dataset size may affect the robustness and external validity of our conclusions. Future work will address these limitations by expanding evaluations across diverse programming languages, incorporating broader semantic and behavioral analyses, and employing larger-scale benchmarks.

\section{Conclusion}

In this paper, we introduced entropy-based metrics—Jensen-Shannon divergence (JSD) and structural cross-entropy (SCE)—to quantify the stability of LLM-generated code using Abstract Syntax Trees (ASTs). Our experiments demonstrate that these metrics:

\begin{enumerate}
\item Provide insights beyond functional correctness (pass@k) and lexical-syntactic metrics (BLEU, CodeBLEU).
\item Correlate strongly with AST-based structural metrics (e.g., TSED), effectively capturing structural stability.
\item Offer significant advantages with structural-only abstraction by reducing token-value noise.
\end{enumerate}

Distinguishing structural-only from full-subtree encodings highlights deep syntactic stability versus token-specific variations. These metrics are lightweight, language-agnostic, and generalizable, complementing existing stability assessments.

Future work will incorporate semantic relationships, such as data- and control-flow dependencies, to extend analysis to behavioral stability. We also plan adaptive subtree-weighting schemes and benchmarking on larger, multi-module codebases to study long-range structural variability.

\section*{Acknowledgement}
The FNR funded this research under grants NCER22/IS/16570468/NCERFT.

\bibliographystyle{IEEEtran}
\bibliography{custom}

\end{document}

%% file: table_compare.tex
\begin{table}[htbp]
\caption{Comparison of current stability metrics with our approaches.}
\label{tab:compare}
\resizebox{\columnwidth}{!}{%
\begin{tabular}{l|l|r|r}
\hline
\begin{tabular}[c]{@{}l@{}}Evaluation\\ Method\end{tabular} &
  Description &
  \multicolumn{1}{l|}{Structure} &
  \multicolumn{1}{l}{Ref.} \\ \hline
\begin{tabular}[c]{@{}l@{}}BLEU/\\ ROUGE-L/\\ METEOR\end{tabular} &
  \begin{tabular}[c]{@{}l@{}}N-gram overlap or sequence matching \\ to reference code.\end{tabular} &
  No & \cite{papineni2002bleu,banerjee-lavie-2005-meteor,lin2004rouge}
   \\ \hline
Exact Match &
  \begin{tabular}[c]{@{}l@{}}Checks if generated code exactly\\ matches the reference.\end{tabular} &
  No & -
   \\ \hline
Pass@k &
  \begin{tabular}[c]{@{}l@{}}Runs up to k generated outputs; \\ succeeds if any pass all tests.\end{tabular} &
  Indirect & \cite{kulal2019spoc}
   \\ \hline
CodeBLEU &
  \begin{tabular}[c]{@{}l@{}}Combines BLEU with AST subtree\\ match and data-flow match.\end{tabular} &
  Yes & \cite{ren2020codebleu}
   \\ \hline
RUBY &
  \begin{tabular}[c]{@{}l@{}}Compare Program Dependency Graph \\ of output vs. reference.\end{tabular} &
  Yes & \cite{paul2024benchmarks}
   \\ \hline
TSED &
  \begin{tabular}[c]{@{}l@{}}Computes the tree edit distance between\\  AST of output and AST of reference.\end{tabular} &
  Yes & \cite{song2024revisiting}
   \\ \hline
\begin{tabular}[c]{@{}l@{}}Semantic\\ Entropy\end{tabular} &
  \begin{tabular}[c]{@{}l@{}}Clusters multiple answers to the same \\ question, then calculates entropy.\end{tabular} &
  Indirect & \cite{ThoughtWorks:2024}
   \\ \hline
\textbf{\begin{tabular}[c]{@{}l@{}}Structural\\ Entropy\tablefootnote{Proposed by this paper. Readers can check the code with the link: \\ \url{https://github.com/Etamin/SCE.git}}\end{tabular}} &
  \begin{tabular}[c]{@{}l@{}}Proposed: Parse the answers' AST structure and \\ computes the subtree entropy between \\ answers, with KL or JS divergence.\end{tabular} &
  Yes & -
  \\ \hline
\end{tabular}%
}\vspace{-2em}
\end{table}

%% file: approaches.tex
\section{Structural Entropy and Similarity}
\SetKwInOut{KwIn}{Input}
\SetKwInOut{KwOut}{Output}
Our pipeline consists of three phases: subtree extraction from ASTs, constructing empirical distributions from these subtrees, and computing similarity metrics. We parse code outputs into ASTs and extract depth-bounded subtrees, transforming them into canonical encodings. These encodings are then used to construct empirical distributions that capture structural variations between code outputs. Finally, we calculate stability using two entropy-based metrics: Jensen–Shannon divergence, providing a symmetric measure of structural similarity, and Structural Cross-Entropy, which emphasizes missing high-probability patterns. 

\subsubsection*{Phase 1\,: Depth-bounded Subtree Extraction}
\label{sec:ph1}
Let $\mathcal{T}_A$ and $\mathcal{T}_B$ denote the ASTs parsed from two code snippets.  Fix a depth parameter $d\in\mathbb{N}$.  For every node $v$ in an AST, we consider the rooted subtree $\mathrm{sub}(v;d)$ which is the fragment of $\mathcal{T}$ rooted at $v$ containing all descendants up to depth $d$.
To transform such subtrees into hashable symbols, we define two canonical encodings:
\begin{enumerate}[label=(\alph*),nosep,leftmargin=2.5em]
  \item \emph{Structure-only\,} encoding  
        $\sigma_{\text{struct}}\!: \mathrm{sub}(v;d)\;\mapsto\;(\text{node‐type}(v),\;
              \bigl(\text{node‐type}(c_1),\dots,\text{node-type}(c_k)\bigr))$,
        where $c_1,\dots,c_k$ are the immediate children of~$v$.
  \item \emph{Structure+value\,} encoding  
        $\sigma_{\text{value}}\!: \mathrm{sub}(v;d)\;\mapsto\;(\text{node-type}(v),\;
              \text{lexeme}(v),\; \\
              \bigl(\text{node-type}(c_1),\dots,\text{node-type}(c_k)\bigr)),
        $
        where $\text{lexeme}(v)$ is the exact source text covered by $v$ (for
        leaves) or a sentinel $\varnothing$ (for internal nodes).
\end{enumerate}
These two approaches trade off generality vs. specificity. Structure-only patterns may capture common coding patterns and yield higher overlap between different programs, but they lose fine-grained information. Structure-with-value patterns are more discriminative (sensitive to exact code), but they also increase the size of the pattern vocabulary and may yield sparser overlap. In practice, one can choose the representation depending on whether value information is important to the similarity task.

\vspace{0.5em}
\noindent
For either choice of $\sigma$, we enumerate the multisets of depth-bounded subtrees extracted from~$\mathcal{T}_A$ and~$\mathcal{T}_B$:
\begin{equation}
\begin{split}
\begin{aligned}
  \label{eq:SA_SB}
  S_A \;&=\;
  \bigl\{\sigma\bigl(\mathrm{sub}(v;d)\bigr)\,\bigm|\,
        v\in \mathcal{T}_A\bigr\},
  \qquad \\
  S_B \;&=\;
  \bigl\{\sigma\bigl(\mathrm{sub}(v;d)\bigr)\,\bigm|\,
        v\in \mathcal{T}_B\bigr\}.
\end{aligned}
\end{split}
\end{equation}
Let $n_A = |S_A|$ and $n_B = |S_B|$ be the total numbers of (possibly repeated) subtree symbols harvested.  In \eqref{eq:SA_SB} we treat $S_A, S_B$ as multisets symbols may appear with multiplicity.

\subsubsection*{Phase 2\,: Constructing Empirical Distributions}
\label{sec:phase2-math}

Let the multisets of subtree symbols extracted in Phase 1 be \(S_A\subseteq\Sigma^{\!*}\) and \(S_B\subseteq\Sigma^{\!*}\), where \(\Sigma^{\!*}\) denotes the countable universe of canonical subtree encodings (either structure–only or structure–with–value). Define the \emph{joint support}
\begin{equation}\label{eq:U}
   U \;=\; S_A \cup S_B
   \;=\;
   \bigl\{u_1,\dots,u_m\bigr\},
   \qquad
   m \;=\; |U|.
\end{equation}
Each \(u_i\in U\) is a distinct subtree symbol.

\paragraph*{Multiplicity functions.}
For \(u\in U\) define
\begin{equation}
\begin{split}
\begin{aligned}
   c_A(u) \;&=\;
   \#\bigl\{\,s\in S_A : s = u\,\bigr\},
   \qquad \\
   c_B(u) \;&=\;
   \#\bigl\{\,s\in S_B : s = u\,\bigr\},
\end{aligned}
\end{split}
\end{equation}
i.e.\ the number of occurrences of \(u\) in the respective multiset. Let
\begin{equation}
\begin{split}
\begin{aligned}
   n_A \;=\; \sum_{u\in U} c_A(u),
   \qquad
   n_B \;=\; \sum_{u\in U} c_B(u)
\end{aligned}
\end{split}
\end{equation}
be the total numbers of (depth-bounded) subtrees harvested from \(\mathcal{T}_A\) and \(\mathcal{T}_B\), respectively.

\paragraph*{Empirical probability distributions.} We define the frequency distributions \(P, Q : U \to [0,1]\) by
\begin{equation}
\begin{split}
\begin{aligned}\label{eq:PQ}
   P(u) \;&=\; \frac{c_A(u)}{n_A},
   \qquad\\
   Q(u) \;&=\;
   \max\!\Bigl(\frac{c_B(u)}{n_B},\,\varepsilon\Bigr),
   \qquad\\
   \sum_{u\in U} P(u) \;&=\;
   \sum_{u\in U} Q(u) \;=\; 1,
\end{aligned}
\end{split}
\end{equation}
where \(0<\varepsilon\ll 1\) is a fixed smoothing constant ensuring \(Q(u)>0\) for every \(u\in U\). (The summation constraint on \(Q\) can be restored by a final renormalisation, but in practice \(\varepsilon\!\ll\!1/n_B\) suffices and the effect on \(\sum_u Q(u)\) is negligible.)

In vector notation let
\begin{equation}
\begin{split}
\begin{aligned}
   \mathbf{p} \;&=\; \bigl(P(u_1),\dots,P(u_m)\bigr)^{\!\mathsf{T}},
   \qquad\\
   \mathbf{q} \;&=\; \bigl(Q(u_1),\dots,Q(u_m)\bigr)^{\!\mathsf{T}}
   \;\in\; [0,1]^m
\end{aligned}
\end{split}
\end{equation}
so \(\lVert\mathbf{p}\rVert_1=\lVert\mathbf{q}\rVert_1=1\). These two probability vectors fully characterize the structural 'vocabularies' of AST \(\mathcal{T}_A\) and \(\mathcal{T}_B\) and constitute the input for the similarity metrics of Phase 3.

\vspace{0.3em}
\noindent
\textbf{Remark.}  
The inclusion of a smoothing parameter \(\varepsilon\) is only required for the directed cross-entropy in Phase 3\,A to avoid undefined \(\log Q(u)\) when \(Q(u)=0\). The symmetric Jensen–Shannon divergence of Phase 3\,B remains finite without explicit smoothing as long as \(P\) and \(Q\) are defined on the common support \(U\).

\subsubsection*{Phase 3\,: Entropy–Based Similarity Metrics}
\label{sec:phase3}

With $P$ and $Q$ defined in~\eqref{eq:PQ}, we present two similarity measures.

\paragraph*{\textbf{(A) Structural Cross-Entropy (SCE)}} Define the cross-entropy of $P$ relative to $Q$ and the Shannon entropy of $Q$ as
\begin{equation}
\begin{split}
\begin{aligned}
   H(P,Q)\;&=\;-\sum_{u\in U}P(u)\log Q(u),\qquad\\
   H(Q)\;&=\;-\sum_{u\in U}Q(u)\log Q(u).
\end{aligned}
\end{split}
\end{equation}
We normalise by taking the \emph{ratio}
\begin{equation}
\begin{split}
\begin{aligned}\label{eq:SCE}
   S_{\mathrm{CE}}(P,Q)\;&=\;
   \frac{H(Q)}{H(P,Q)},\qquad 0<S_{\mathrm{CE}}\le 1. \\
   S_{\mathrm{CE}}(P,Q)&=1
   \ \Longleftrightarrow\ 
   P=Q;\quad \\
   S_{\mathrm{CE}}&\to 0
   \ \text{as}\ Q\ \text{fails to explain}\ P.
\end{aligned}
\end{split}
\end{equation}
Because the numerator and denominator share the same logarithmic base, the score is scale–independent and monotone in the KL divergence $D_{\mathrm{KL}}(P\parallel Q)=H(P,Q)-H(P)$. In code, compute $H(Q)$ and $H(P,Q)$ directly with base-2 logs; skip terms with $P(u)=0$ and apply $\varepsilon$ smoothing on $Q(u)$. This ratio acts as a normalized \textbf{structural cross-entropy}.

\paragraph*{\textbf{(B) Jensen–Shannon Divergence (JSD)}} Let $M=\tfrac12(P+Q)$ denote the midpoint distribution. The Jensen–Shannon divergence is
\begin{equation}
\begin{split}
\begin{aligned}
   D_{\mathrm{JS}}(P\parallel Q)
   \;&=\;
   \tfrac12D_{\mathrm{KL}}(P\parallel M)
   \;+\;
   \tfrac12D_{\mathrm{KL}}(Q\parallel M)\\
   \;&=\;
   H(M)-\tfrac12\!\bigl[H(P)+H(Q)\bigr].
\end{aligned}
\end{split}
\end{equation}
With logarithm base 2, $D_{\mathrm{JS}}\!\in[0,1]$\,bits and $\sqrt{D_{\mathrm{JS}}}$ is a metric \cite{lin2002divergence}.
We use
\begin{equation}\label{eq:SJSD}
   S_{\mathrm{JSD}}(P,Q)\;=\;1-D_{\mathrm{JS}}(P\parallel Q)\in[0,1],
\end{equation}
where larger values again indicate greater similarity. Unlike $S_{\mathrm{CE}}$, JSD is \emph{symmetric} and finite without smoothing so long as $P,Q$ share support.

\paragraph{Computational cost}
Both scores require $O(|U|)$ arithmetic once $P$ and $Q$ are formed. When depth $d$ is fixed, $|U|$ grows at most linearly in the number of AST nodes, the overall pipeline $O(n\,d)$ for subtree extraction plus $O(|U|)$ for scoring. 

\paragraph{Summary}
$S_{\mathrm{CE}}$ (Equation~\ref{eq:SCE}) offers a directed measure—“how efficiently does distribution $Q$ encode $P$?”—while $S_{\mathrm{JS}}$ (Equation~\ref{eq:SJSD}) gives a symmetric, bounded divergence–to–similarity conversion. Either can be deployed depending on whether directionality or symmetry is desired; both map naturally into the $[0,1]$ range, facilitating thresholding and comparative analysis of program structures.

%% file: exp.tex
\section{Experiments and Results}
\input{result_table}
To evaluate the effectiveness and interpretability of both entropy-based stability metrics we proposed, (1) Jensen–Shannon Divergence (JSD) and (2) Structural Cross Entropy (SCE), we conducted experiments focusing on the stability of LLM-generated code. Specifically, we provided each prompt to the LLM five times independently, generating five distinct outputs per prompt. We then measured stability by computing similarity or divergence scores pairwise among these five outputs, using standard evaluation metrics (BLEU, CodeBLEU, TSED) and our proposed entropy-based metrics (JSD and SCE). The final stability score for each metric was obtained by averaging all pairwise comparisons.

For comparison, we also included pass@k, a metric evaluating functional correctness, computed by executing each generated code output against provided test cases. Unlike other metrics, pass@k directly reflects correctness rather than stability but serves as an important baseline for understanding practical implications of structural stability. Additionally, we analyzed structural-only variants -- JSD (structural) and SCE (structural) -- (cf. Section \ref{sec:ph1}.1(a)) by abstracting away token values, thus isolating structural differences to examine their effect on stability.

\subsection{Experimental Setup}

We performed our evaluations using two widely recognized benchmarks across two programming languages:

\begin{itemize}
\item \textbf{Python (BigCodeBench):} A dataset widely adopted for general code generation tasks~\cite{zhuo2024bigcodebench}.
\item \textbf{SQL (Spider):} A comprehensive benchmark for SQL query generation~\cite{yu2018spider}.
\end{itemize}

We evaluated three representative models: LLaMA 3.1 (8B, instruct), Qwen-2.5(7B, instruct), and Qwen-2.5-Coder(7B). Our analysis focuses on understanding how our entropy-based measures reflect code stability and their correlations with existing metrics.

\subsection{Quantitative Results}
\begin{figure}
    \vspace{-1em}
    \centering
    \includegraphics[width=0.9\linewidth]{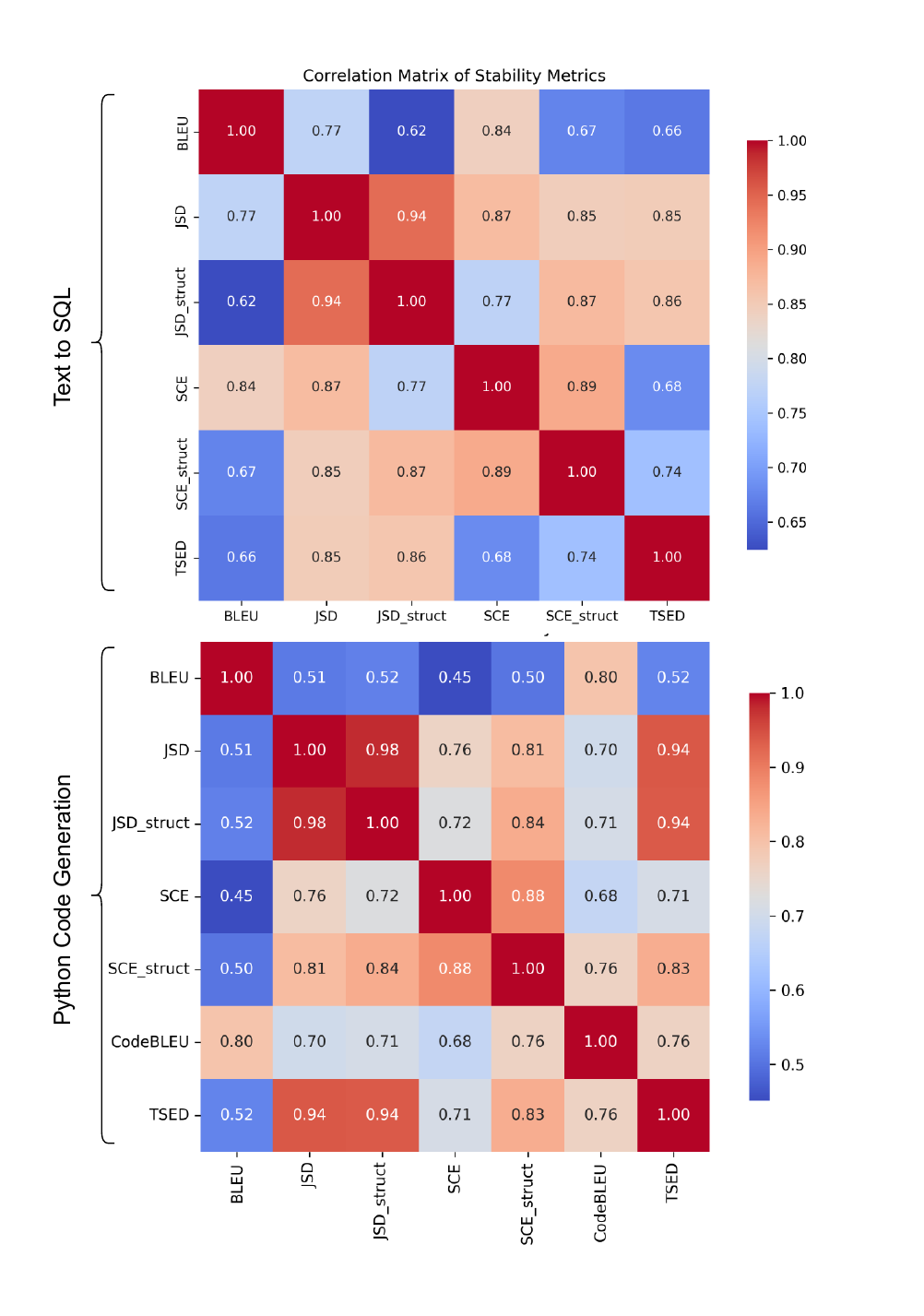}
    
    \caption{Heatmap of Pearson Correlation Coefficient between Stability Evaluation Metrics, JSD: Jensen-Shannon Divergence; SCE: Structural Cross Entropy, Structural means value in subtree is ignored.}
    \label{fig:heatmap}
    \vspace{-1em}
\end{figure}
The results are summarized in Table II. For the Python tasks, while the Qwen-2.5 models exhibit stronger stability in lexical metrics such as BLEU and CodeBLEU compared to LLaMA 3.1, a notable discrepancy between their pass@1 and pass@5 indicates considerable variability. Specifically, Qwen-2.5-Coder achieves pass@1 of 0.373 and pass@5 of 0.517, suggesting that multiple sampled outputs differ substantially in correctness, reflecting inherent instability. 

For SQL (Spider), this instability is also evident. Although Qwen models achieve higher overall correctness and stability, they demonstrate substantial differences between pass@1 and pass@5 (e.g., Qwen-2.5-Coder: pass@1 = 0.787, pass@5 = 0.860), further highlighting output variability.

Our proposed entropy-based stability metrics (JSD and SCE) complement these findings. JSD values remain consistently high (above 0.9), suggesting that structurally, outputs are broadly similar across samplings. However, SCE scores (especially when considering token values explicitly) are consistently lower, indicating sensitivity to subtle variations overlooked by purely structural metrics. This contrast highlights the utility of SCE in detecting fine-grained token-level variability.

Moreover, the structural-only variants of both metrics generally produce higher scores, as expected, reflecting their insensitivity to identifier or literal changes. Thus, these structural-only scores represent a baseline stability that isolates deeper syntactic patterns from superficial token-level differences.

Overall, our metrics successfully quantify and clarify the nature of stability in generated code, complementing traditional correctness-based metrics like pass@k, which indirectly reflect variability through the disparity between single- and multi-sample correctness evaluations.

\subsection{Correlation Analysis}

Figure \ref{fig:heatmap} illustrates the Pearson correlation coefficient across stability metrics:

\paragraph{Python (BigCodeBench)} We observe high correlations between JSD and JSD(structural) (0.94), indicating structural abstraction maintains strong stability signals. The correlation between SCE and JSD metrics is moderately strong (0.85-0.87), though lower than the near-perfect correlation between JSD and its structural variant (0.94). TSED shows good correlation with JSD (0.85–0.86), reinforcing its utility for capturing structural differences. Notably, SCE structural variants correlate less strongly with TSED (0.74), highlighting differences in sensitivity to structural variations.

\paragraph{SQL (Spider)} On SQL tasks, the correlations between JSD and JSD(structural) are even stronger (0.98), underscoring the robustness of structural abstraction in capturing stability. Interestingly, BLEU's correlation is lower with JSD metrics (around 0.51), emphasizing the limitation of token-based metrics in SQL's structural context. CodeBLEU is not valid for SQL, hence excluded. TSED correlates strongly with JSD (0.94), confirming their structural similarity evaluation alignment. The SCE metrics correlate moderately (0.72–0.88) with others, again highlighting their distinctive sensitivity to token-level discrepancies.

%% file: result_table.tex
\begin{table*}[htbp]
\caption{Comparison of current stability metrics. JSD: Jensen-Shannon Divergence; SCE: Structural Cross Entropy; structural means only compares the subtree structure.}
\label{tab:result}
\resizebox{2\columnwidth}{!}{%
\begin{tabular}{llrrrrrrrrr}
\hline
Model &
  Language(Task) &
  \multicolumn{1}{l}{Avg. BLEU} &
  \multicolumn{1}{l}{Code BLEU} &
  \multicolumn{1}{l}{Pass@1} &
  \multicolumn{1}{l}{Pass@5} &
  \multicolumn{1}{l}{TSED} &
  \multicolumn{1}{l}{SCE(structural)} &
  \multicolumn{1}{l}{SCE} &
  \multicolumn{1}{l}{JSD(structural)} &
  \multicolumn{1}{l}{JSD} \\ \hline
LLaMA-3.1 8B it & \multirow{3}{*}{\begin{tabular}[c]{@{}l@{}}Python\\ BigCodeBench\end{tabular}} & 0.428 & 0.669 & 0.289 & 0.481 & 0.785 & 0.798 & 0.656 & 0.940 & 0.898 \\
Qwen-2.5 7B it        &                                                                                & 0.596 & 0.716 & 0.339 & 0.482 & 0.765 & 0.823 & 0.726 & 0.947 & 0.913 \\
Qwen-2.5-Coder 7B &                                                                                & 0.614 & 0.715 & 0.373 & 0.517 & 0.764 & 0.823 & 0.722 & 0.951 & 0.918 \\ \hline
LLaMA-3.1 8B it & \multirow{3}{*}{\begin{tabular}[c]{@{}l@{}}SQL\\ Spider\end{tabular}}          & 0.495 & N/A   & 0.673 & 0.824 & 0.832 & 0.729 & 0.669 & 0.934 & 0.905 \\
Qwen-2.5 7B it       &                                                                                & 0.770 & N/A   & 0.716 & 0.790 & 0.921 & 0.886 & 0.850 & 0.971 & 0.958 \\
Qwen-2.5-Coder 7B &                                                                                & 0.664 & N/A   & 0.787 & 0.860 & 0.927 & 0.822 & 0.781 & 0.962 & 0.946 \\ \hline
\end{tabular}%
}\vspace{-1em}
\end{table*}